\documentstyle[prl,aps,epsfig,twocolumn]{revtex}

\draft

\newlength{\textwidthm}
\begin{document}
\title{Atom correlations and spin squeezing near the Heisenberg limit:
finite size effect and decoherence}
\author{A. Andr\'e and M.D. Lukin}
\address{Physics Department, Harvard University and \\ 
ITAMP, Harvard-Smithsonian Center for Astrophysics, Cambridge, MA~~02138}
\date{\today}
\maketitle
\begin{abstract}

We analyze a model for spin squeezing based on the so-called counter-twisting
Hamiltonian, including the effects of dissipation and finite system size. 
We discuss the conditions under which the Heisenberg limit, i.e. phase 
sensitivity $\propto 1/N$, can be achieved.
A specific implementation of this model based on atom-atom
interactions via quantized photon exchange is presented in detail.
The resulting
excitation corresponds to the creation of spin-flipped atomic pairs
and can be used for
fast generation of entangled  atomic ensembles, spin squeezing and
applications in quantum information processing.
The conditions for achieving strong spin squeezing with this mechanism are
also analyzed.

\end{abstract}



\section{Introduction}

Interacting quantum systems that start in uncorrelated states generally
evolve towards entangled states due to quantum correlations building up in
time. These correlations and the form they take depend crucially on the
interaction that gives rise to them. For example in parametric down
conversion or in the optical parametric oscillator (OPO) pairs of photons
can be created in distinct modes of the electromagnetic field. The fact
that {\it pairs} of photons are generated leads to quantum correlations
between the two modes. Since each mode is described by a harmonic
oscillator, one can think of the state of the field as the quantum state
of two fictitious particles in harmonic oscillator potentials. The
quantum correlations correspond to e.g. the positions of the particles
being strongly correlated, in the ideal case $\Delta
(X_1-X_2)^2\rightarrow 0$ and their
momenta being anticorrelated $\Delta (P_1+P_2)^2\rightarrow 0$. For the
electromagnetic field modes, the position and momenta correspond to
quadratures of the field modes and it is between these that correlations
are produced \cite{Kimble-Houches,Walls-Milburn}.
These correlations
are essential to quantum communication e.g. quantum teleportation of 
information from one location to another \cite{Q-teleport}. 
Entanglement is also crucial for many schemes in quantum cryptography and
for long-distance quantum communication through lossy channels 
\cite{Longdist}. 

Since the mechanism for producing correlations in electromagnetic field
modes is at the fundamental level so simple (photons created in pairs) it
is natural to wonder if such a simple mechanism may lead to entanglement
of atoms interacting in a similar manner. In complete analogy to the OPA
mechanism, a process that transfers {\it pairs} of atoms from their ground
state to two well defined final states also gives rise to quantum
correlations between atoms. When a collection of $N$ two level atoms is
thought of as an ensemble of effective spin ${\tiny \frac{1}{2}}$
particles with total pseudo-angular momentum $J=N/2$, it turns out
\cite{Ueda} that the quantum correlations produced by an interaction that
transfers atoms in pairs from the lower state to the upper state shows up
as reduced fluctuations in a component of the angular momentum e.g.
$\Delta J_x^2 \rightarrow 0$. 
We will discuss entanglement of atoms with
one another in an atomic ensemble for which an effective interaction
leads to the transfer of atoms in pairs to well defined final states and
we will use the concept of spin squeezing to quantify the amount of
quantum correlations produced in such a case. As for squeezed states of
light, decoherence mechanisms and dissipation are acting in such a way as
to destroy or limit the amount of squeezing achievable in practice. We
also analyze the influence of such dissipation mechanisms and
find relations between the spin squeezing interaction rate, the
dissipation rate and the amount of squeezing achievable in the presence of
damping mechanisms.
The coherent control of the dynamical evolution of complex systems such as
atomic ensembles may lead to the
production of entangled non-classical states such as spin squeezed states
\cite{Ueda} (analogous to squeezed states of light \cite{Walls-Milburn})
and correlated collective atomic modes (similar to twin
photons generated by a non-degenerate OPO). 

The main result of this paper is that for a collection of $N$ atoms with
average single atom nonlinearity $\chi$ (two atom interaction energy) and
with single atom loss rate $\Gamma$, the condition for achieving some
spin squeezing is that $N\chi > \Gamma$.
In order to achieve reduction of uncertainty in say $J_x$ compared to the
uncertainty in the Bloch state $|J=N/2,J_z=N/2\rangle$ for which $(\Delta
J_x)^2=N/4$ by an amount $s$ (i.e. $(\Delta J_x)^2=N/(4s)$)
with $1\leq s \leq N$, one requires that $N\chi>s\Gamma$ and the
interaction time needed scales as $t\sim (\log{s})/(N\chi)$ while the
maximum number of atoms than can be lost without destroying the squeezing
scales as $\Delta N\sim (N/s)\log{s}$.
To achieve Heisenberg limited precision (i.e. maximum
spin squeezing $s\sim N$), one needs a large single atom nonlinearity
$\chi>\Gamma$. This means that the interaction time needed to achieve this
strongly correlated state is $t\sim (\log{N})/(N\chi)$ and the maximum
number of atoms that can be
lost without compromising this optimal squeezing is $\Delta N\sim
\log{N}$ i.e. a very small number of atoms lost may prevent reaching the
Heisenberg limit. This analysis remains valid and agrees with a specific
implementation
based on an effective atom-atom interaction via quantized photon
exchange in a cavity, for which the decoherence mechanism corresponds to
spontaneous emission and leakage of photons from the cavity.  

The possibility of coherently controlling interacting quantum systems
has lead to many new developments in the field of quantum information 
science \cite{Bouwmeester}.
These are expected to have an impact in a broad area ranging from quantum
computation and quantum communication \cite{Nielsen-Chuang} to precision
measurements \cite{Wineland} and controlled modeling of complex quantum
phenomena \cite{Sachdev}.
Entangled systems realized in the laboratory range in size from few qbits
\cite{Wineland2}, to macroscopic ensemble of particles \cite{Polzik}.
Controllable coherent interactions between atoms \cite{Lloyd,Meystre} may
also open the 
way for modelling of complex quantum phenomena such as quantum phase 
transitions \cite{Sachdev} in which quantum correlations play a crucial role.

Entanglement of a single atomic ensemble, i.e. quantum correlations between
atoms in the same ensemble, has been shown to be potentially very 
useful in the field of precision measurements \cite{Wineland}. 
Certain types of interactions between atoms lead to 
entanglement and spin squeezing, 
characterized by reduced variance in an
observable and increased fluctuations in the canonically conjugate observable.
This reduction of fluctuations directly translates into an improved
accuracy for
measurements sensitive to that observable. A typical figure of merit for spin
squeezed states is the phase accuracy $\delta\phi$ on 
estimating accumulated dynamical phase in the Ramsey interferometric 
experiment. With all experimental uncertainties controlled below this noise 
level, the dominant source of noise in such experiments is the ``quantum 
projection noise'' \cite{Wineland} associated with e.g. the noise in
measurements of the 
x-component of the spin of an ensemble of two-level atoms (effective spin 
${\tiny \frac{1}{2}}$) all
prepared in the lower level (the $|\downarrow\rangle$ state). This noise leads
to a lower limit on phase accuracy $\delta\phi=1/\sqrt{N}$ called the 
standard quantum limit (SQL), where $N$ is 
the number of atoms in the ensemble. 
The Heisenberg uncertainty principle 
however allows for phase accuracies consistent with the basic principles of
quantum mechanics that are as low as $\delta\phi=1/N$, called the
Heisenberg limit.

We also discuss in more detail a technique \cite{Paper1} based on a
resonantly enhanced
nonlinear process involving Raman scattering into a 
``slow'' optical mode \cite{slow-light}, which creates a 
pair of spin-flipped atom and slowly propagating coupled excitation
of light and matter (dark-state  polariton). When the group
velocity of the polariton is reduced to zero \cite{stop-light,darkpolar}, 
this results in pairs of spin flipped atoms. 
The dark-state polariton can be easily converted into  corresponding states 
of photon wavepackets ``on demand'' \cite{darkpolar}, 
which makes  the present approach most suitable for implementing protocols 
in quantum information  processing that require a combination of 
deterministic sources of entangled states and 
long-lived quantum memory \cite{Longdist,Q-mem}.

This paper is divided into V sections. In Section II, we discuss Ramsey 
spectroscopy and the use of spin-squeezed states in precision
measurements.
In particular we analyze the situation where $N$ two level atoms with levels
$|g\rangle$ and $|e\rangle$ are prepared in a correlated state and 
subsequently
probed by separated fields of frequency $\omega$ in the Ramsey
interferometric configuration, which we review in Appendix A.
We also discuss spin-squeezed states and develop pictorial
representation of those states which we compare to squeezed states of
light and in appendix B we introduce the Wigner representation for a
particular class of spin squezed states.

In Section III, we analyze a model for spin squeezing based on the analogy
with 
the optical parametric oscillator. We also seek to understand the influence of 
loss processes on the coherent spin-squeezing interaction and the way in which 
it limits the correlations achievable for a given interaction rate. The
model 
consists of two bosonic modes (a ``spin up'' state and a ``spin down'' state) 
with loss
rates and a coherent interaction that transfers pairs of atoms from one mode
to the other.
For our simple model, analytical results can be obtained
in the perturbative regime of small number of excitations (most atoms in the
lower state) and low loss rate. We estimate the
conditions for which Heisenberg-limited spin squeezed states can be
produced.

In Section IV, we present a scheme for inducing effective coherent
interactions 
between atoms in an atomic ensemble. These coherent interactions 
lead to massive entanglement of the ensemble and to characterize the degree 
of entanglement thus obtained, we calculate the squeezing or reduction in 
fluctuations of one particular observable. The coherent interaction is based
on Raman scattering into a cavity mode for which the atomic medium is made 
transparent by Electromagnetically Induced Transparency (EIT). The slowly
propagating mode is then best described by a polariton: a collective
excitation that is partly photonic and partly ``spin'' excitation of the atomic
ensemble (the up and down states of the spin being two metastable states). 
The overall process leads to the creation of pairs of
excitations, one being a ``spin flip'' created by Raman scattering, the other 
being a polariton which can be ``steered'' into a photon or spin flip
excitation ``on demand''. We find that substantial spin squeezing can be 
obtained for atomic ensembles in low finesse cavities, without the
strong coupling requirement
of cavity QED. In the limit of unity finesse this corresponds to free-space 
configuration and substantial correlations can still be produced in this
case. In the 
opposite limit of high finesse, very strong correlations are obtained and 
in particular we estimate the regime for which Heisenberg limited spin-squeezed
states are produced.

\section{Ramsey spectroscopy with correlated atoms}

In appendix A, Ramsey spectroscopy is reviewed and in particular 
we show how
the phase accuracy in phase estimation based on the Ramsey fringe signal is,
at the maximum sensitivity point, given by

\begin{equation}
\delta\phi=\frac{\Delta J_x}{|\langle \hat{J}_z \rangle|}
\end{equation}

where $\Delta J_x$ is the variance in the x-component of the pseudo angular 
momentum (of length $J=N/2$)
representing the state of $N$ two-level atoms and $\langle \hat{J}_z \rangle$
is the expectation value of the z-component of the pseudo angular momentum
(both the variance and the expectation value are
calculated in the initial state). 

For an uncorrelated state of atoms e.g. with all atoms in their lower
state so that the state of the ensemble is described by
$|J_z=-N/2\rangle$, it is found that 
$\Delta J_x=\sqrt{J/2}$ and $\langle \hat{J}_z \rangle=-J$ so that 
$\delta\phi=\sqrt{1/N}$.
In order to improve the phase accuracy, one must use
a state for which the variance in
$\hat{J}_x$ is reduced while $\langle \hat{J}_z \rangle$ is little
changed. 
Consider therefore a state such as an 
eigenstate of $\hat{J}_x$, for example $|J_x=0\rangle$. Calculating the
expectation values and variances we find $\langle
\hat{J}_x \rangle = 
\langle \hat{J}_y \rangle = \langle \hat{J}_z \rangle = 0$, $\Delta J_x = 0$ 
and $\Delta J_y = \Delta J_z = \sqrt{J(J+1)/2}$. However the Ramsey signal
has amplitude proportional to $\langle J_z \rangle$ and therefore vanishes
for all phase angles $\phi$, which means that even though the noise or
fluctuation properties of the signal may be improved, its average is zero.
Note that this is
because we have chosen $\hat{J}_z(\phi)$ as our observable, other
observables such as $\hat{J}_z^2(\phi)$ for example may lead to non-zero 
average signal together with reduced variance
\cite{Kasevich,Holland}. However, it turns out their signal to noise ratio
is very much reduced compared to that of the Ramsey scheme \cite{Holland}. 
It is thus necessary to consider
states that lead to a reduced variance $\Delta J_x$ while maintaining a large 
signal amplitude, i.e. a large $\langle \hat{J}_z \rangle$. We therefore 
consider states such as 

\begin{equation}
|\psi(a)\rangle = \frac{1}{\sqrt{1+a^2}}(i|J_x=0\rangle + 
a\frac{|J_x=+1\rangle - |J_x=-1\rangle}{\sqrt{2}})
\label{psia}
\end{equation}

where $a$ is a real number parametrizing the state $|\psi(a)\rangle$.
It is straightforward to calculate the expectation values 
\begin{eqnarray}
\langle \hat{J}_x \rangle &=& 0 \nonumber \\
\langle \hat{J}_y \rangle &=& 0 \nonumber \\
\langle \hat{J}_z \rangle &=& \frac{2a}{1+a^2} \sqrt{\frac{J(J+1)}{2}}
\end{eqnarray}

and the variances 
\begin{eqnarray}
\Delta J_x &=& \frac{a}{\sqrt{1+a^2}} \nonumber \\
\Delta J_y &=& \frac{1}{\sqrt{1+a^2}}\sqrt{\frac{J(J+1)}{2}} \nonumber \\
\Delta J_z &=& \sqrt{\frac{J(J+1)}{2}}\left[1-\frac{4a^2}{(1+a^2)^2}\right]
^{1/2} .
\label{varpsia}
\end{eqnarray}


\begin{figure}[ht]
\begin{center}
\epsfig{file=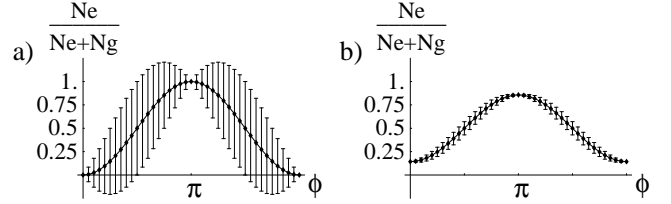,width=8.5cm}
\leavevmode
\end{center}
\vspace*{2ex}
\caption{Number of atoms detected in the upper state ($N_e$) relative
to total number of atoms (the total number is $N=N_e+N_g$
and thus $N_e/(N_e+N_g)=(N/2+\langle J_z(\phi) \rangle )/N$) 
vs. accumulated phase 
$\phi=(\omega-\omega_0)T$ for a) uncorrelated atoms and b) correlated atoms in
a spin-squeezed state $|\psi(a)\rangle$, for $N=100$ and $a=-1.1$ (note
that error bars have been magnified by a factor of $10$ for clarity). 
Note how squeezing of the variance improves the phase accuracy.
}
\end{figure}


The signal amplitude which depends on $\langle \hat{J}_z \rangle$ can thus
be rather large ($O[N]$) while the noise amplitude characterized by 
$\Delta J_x$ is minimized ($O[1]$). 
The Ramsey signal and phase accuracy for such a state is shown in Fig. 1b,
compared 
to the case of uncorrelated atoms (Fig. 1a).

These states are minimum uncertainty states i.e. 
$(\Delta J_x)(\Delta J_y)={\tiny \frac{1}{2}}|\langle \hat{J}_z \rangle |$ 
for all values of the parameter $a$. Also, their phase accuracy is given by 

\begin{equation}
\delta\phi(\pm\pi/2) = \sqrt{\frac{1+a^2}{2}}\frac{1}{\sqrt{J(J+1)}} 
\end{equation}

which is of order $1/N$.
The best phase accuracy is obtained for $a\rightarrow 0$ in which case the
optimal phase accuracy is $\delta\phi(\pm\pi/2)=\sqrt{2}/N$. Note that in
this case the signal amplitude ($\propto \langle\hat{J}_z\rangle$) becomes
vanishingly small ($\langle\hat{J}_z(\phi)\rangle\rightarrow 0$ for all
$\phi$) and also the
range of values of $\phi$ for which improved phase accuracy is achieved
becomes vanishingly small around $\phi=\pm\pi/2$. For these reasons, the
optimally spin squeezed state $|\psi(a=0)\rangle$ may prove impractical.
Note however that for finite $a$ i.e. for $|a|=1$, the signal amplitude is
large ($\sim N/\sqrt{8}$) and 
the phase accuracy is independent of $\phi$

\begin{equation}
\delta\phi(\phi)=\frac{1}{\sqrt{J(J+1)}}\simeq\frac{2}{N}
\end{equation}

i.e. twice the Heisenberg limit. 


\begin{figure}[ht]
\begin{center}
\epsfig{file=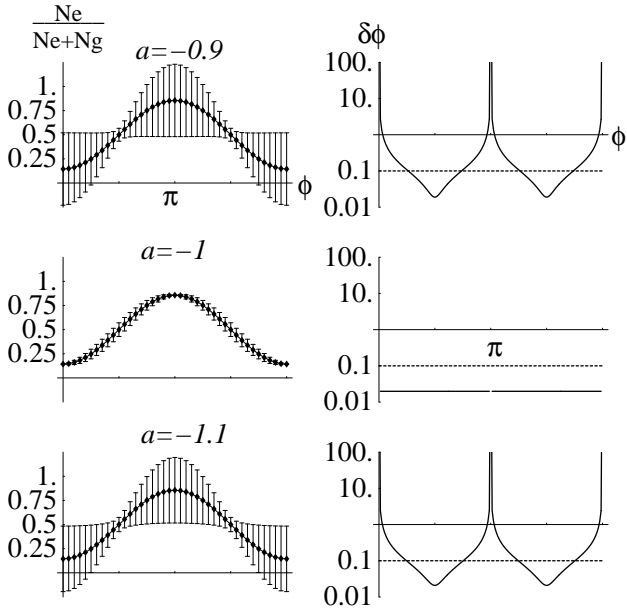,width=8.5cm}
\leavevmode
\end{center}
\vspace*{2ex}
\caption{Number of atoms detected in the upper state vs. accumulated phase 
$\phi=(\omega-\omega_0)T$ for correlated atoms in various spin-squeezed states
$|\psi(a)\rangle$, for $N=100$ and $a=-0.9$, $a=-1$ and $a=-1.1$. 
Also shown is the phase accuracy $\delta\phi(\phi)$ vs. accumulated phase
(the dashed line represents the standard quantum limit
$\delta\phi=1/\sqrt{N}$). 
Note how $\delta\phi(\phi)$ gets to a minimum value of order $2/N=0.02$.
}
\end{figure}


In Fig. 2 we show the signal and variance for various spin squeezed states 
along with the phase accuracy $\delta\phi(\phi)$.

We can gain a better understanding of the squeezing in the states 
(\ref{psia})
$|\psi(a)\rangle$ by looking at various representations of them. 
The simplest representation is to project the state onto eigenstates of the
three components of the angular momentum

\begin{equation}
P_i(m) = |\langle J_i=m|\psi(a)\rangle |^2
\end{equation}

where $|J_i=m\rangle$ is the eigenstate of the i-component of angular momentum
with eigenvalue $m$. 

From Fig. 3 it is clear that the expectation value of $\hat{J}_x$ and
$\hat{J}_y$ are zero in such a state, whereas (for $a=-1$) the expectation 
value of $\hat{J}_z$ is large and negative, the variances are clearly given by
(\ref{varpsia}). It is interesting to note the similarity of these angular
momentum squeezed states and those of a harmonic oscillator (i.e. 
squeezed states of light). In both cases, the probability distributions vanish
for odd number of quanta (note that for simplicity we consider only $N$ even 
here). 


\begin{figure}[ht]
\begin{center}
\epsfig{file=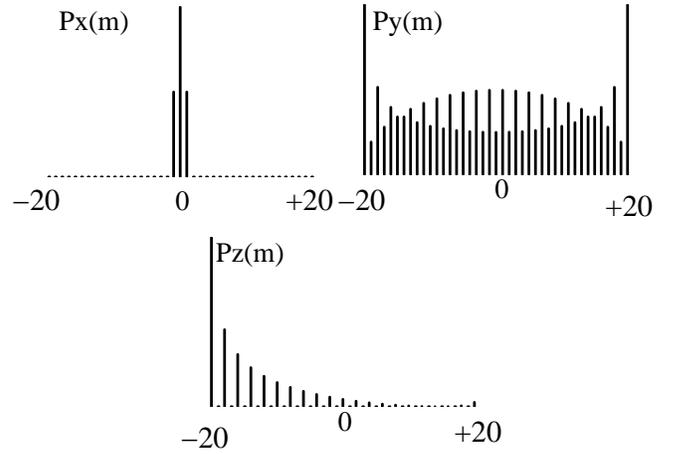,width=8.5cm}
\leavevmode
\end{center}
\vspace*{2ex}
\caption{Projection of the state $|\psi(a)\rangle$ onto eigenstates of the
angular momentum operators $\hat{J}_x$, $\hat{J}_y$ and $\hat{J}_z$ for 
$J=N/2=20$ and $a=-1$. The $P_x(m)$ distribution is sharply peaked since
the 
state $|\psi(a)\rangle$ is a superposition of the $m_x=-1,0,+1$ components
only;
the $P_y(m)$ distribution is broad and symmetric; the $P_z(m)$ distribution
vanishes for $m$ odd and the even components decrease roughly exponentially
with $m$.} 
\end{figure}


For even number of quanta, the behaviour is 
nearly exponential $P_z(m)\propto e^{-c(m+j)}$ for some constant $c$.
A mechanism for generating such states starting from the uncorrelated state 
$|J_z=-J\rangle$ must therefore be one in which atoms are excited in pairs, 
i.e.
two atoms in the ground state are transferred to the excited state 
$|g\rangle |g\rangle \rightarrow |e\rangle|e\rangle$.
Consider the similarities with squeezed states of light: in particular the
photon number distribution vanishes 
for odd photon number in the case of squeezed vacuum due to the form of the
squeezing Hamiltonian $\hat{H}=-i\chi [\hat{a}^{\dagger 2}-\hat{a}^2]$, which 
creates and destroys photons in pairs. Since we find a similar cancellation
of the probability of there being odd number of excitations for the states 
$|\psi(a)\rangle$, the interaction giving rise to such states starting from
all atoms in their lower states must likewise create and destroy excitations 
in pairs and thus be of the form 
$\hat{H}=-i\hbar\chi[\hat{L}_+^2-\hat{L}_-^2]$.
This process can 
also be viewed as a coherent collision mechanism. Moreover, for the whole
atomic ensemble to become entangled (not just particular atom pairs), this
process must occur completely symmetrically for all atoms. It should
not be two particular atoms that get transferred to the excited state, 
rather it should be two collective excitations that get created

\begin{eqnarray}
|g_1\cdots g_N\rangle & \rightarrow & \sqrt{\frac{2}{N(N-1)}}\sum_{i>j}\left[
|g\cdots e_i \cdots e_j \cdots g\rangle\right]
\end{eqnarray}

which is the state obtained by letting $\hat{J}_+^2$ operate on 
$|J_z=-J\rangle$. The simplest Hamiltonian giving rise to this type of
interaction is analyzed in section III to quantify the squeezing generated by 
this mechanism.
This form of interaction was considered by Kitegawa and Ueda \cite{Ueda} in 
their classic study of spin squeezing and was dubbed the ``two axis 
countertwisting'' interaction.

\section{Two axis counter twisting model}

We now turn to the analysis of the two axis countertwisting 
Hamiltonian \cite{Ueda}

\begin{eqnarray}
\hat{H} &=& -i\frac{\hbar\chi}{2}\left( \hat{L}_+^2-\hat{L}_-^2 \right)
\nonumber \\
&=& \hbar\chi \left(\hat{L}_x\hat{L}_y+\hat{L}_y\hat{L}_x\right).
\label{ctham}
\end{eqnarray}

As argued in last section it is this type of Hamiltonian that most 
closely parallels squeezed state generation for light. 

We now present a general theory that allows to quantify atom correlations
and takes into account decoherence and finite system size.
Specifically, we consider two bosonic modes (such as for a two 
component Bose-Einstein condensate or for an atomic ensemble with two relevant
atomic levels) with annihilation 
operators $\hat{a}_1$ 
and $\hat{a}_2$. The system is also subject to damping i.e. loss of atoms at 
rates which may depend on the internal state. 
The equations of motion for the two modes are
then (with $\hat{L}_+=\hat{a}_2^\dagger\hat{a}_1$ and 
$\hat{L}_-=\hat{a}_1^\dagger\hat{a}_2$)

\begin{eqnarray}
\dot{\hat{a}}_1 = -\gamma_1 \hat{a}_1 + \chi\hat{a}_1^\dagger\hat{a}_2^2 + 
\hat{F}_1(t) \nonumber \\
\dot{\hat{a}}_2 = -\gamma_2 \hat{a}_2 - \chi\hat{a}_2^\dagger\hat{a}_1^2 + 
\hat{F}_2(t)
\end{eqnarray}

where $\hat{F}_j(t)$ are delta-correlated Langevin noise forces with 
appropriate diffusion coefficients 
$D_{ij}=\langle \hat{F}_i(t)\hat{F}_j(t) \rangle$.

In order to discuss spin squeezing, it is easier to rewrite the equations of
motion in terms of the Stokes parameters

\begin{eqnarray}
\hat{L}_0 &=& \hat{N} = \hat{a}_1^\dagger\hat{a}_1+\hat{a}_2^\dagger\hat{a}_2 \nonumber \\
\hat{L}_x &=& (\hat{a}_2^\dagger\hat{a}_1+\hat{a}_1^\dagger\hat{a}_2)/2 \nonumber \\
\hat{L}_y &=& (\hat{a}_2^\dagger\hat{a}_1-\hat{a}_1^\dagger\hat{a}_2)/2i \nonumber \\
\hat{L}_z &=& (\hat{a}_2^\dagger\hat{a}_2-\hat{a}_1^\dagger\hat{a}_1)/2
\end{eqnarray}
for which the equations are

\begin{eqnarray}
\dot{\hat{L}}_0 &=& -2\Gamma\hat{L}_0+4\gamma\hat{L}_z+\hat{F}_0(t) 
\nonumber \\
\dot{\hat{L}}_x &=& -2\Gamma\hat{L}_x+\chi(\hat{L}_x\hat{L}_z+
\hat{L}_z\hat{L}_x)+\hat{F}_x(t) \nonumber \\
\dot{\hat{L}}_y &=& -2\Gamma\hat{L}_y-\chi(\hat{L}_y\hat{L}_z+
\hat{L}_z\hat{L}_y)+\hat{F}_y(t) \nonumber \\
\dot{\hat{L}}_z &=& -2\Gamma\hat{L}_z+\gamma\hat{L}_0-2\chi(\hat{L}_x^2-
\hat{L}_y^2)+\hat{F}_z(t) 
\label{fullqmeqns}
\end{eqnarray}

where $\Gamma=(\gamma_1+\gamma_2)/2$, $\gamma=(\gamma_1-\gamma_2)/2$ and 
$\hat{F}_j(t)$ are new delta-correlated noise forces associated with the 
damping.

Since (\ref{fullqmeqns}) are non-linear operator equations,
in the equations of motion for the first order moments 
$\langle \hat{L}_i \rangle$ there are terms that depend on those first order
moments but also terms depending on the second order moments  
$\langle \hat{L}_i\hat{L}_j \rangle$. Similarly the equations of
motion for the second order
moments depend on themselves and also on the third order moments, and so
on, leading to
the BBGKY hierarchy of equations of motion for the expectation values of 
operator products.
In order to solve this set of equations, the hierarchy must be truncated
at some order \cite{Vardi}. Keeping the first and second order
moments, 
we truncate the BBGKY hierarchy by the approximation

\begin{eqnarray}
\langle\hat{L}_i\hat{L}_j\hat{L}_k\rangle & \approx & 
\langle\hat{L}_i\hat{L}_j\rangle\langle\hat{L}_k\rangle +
\langle\hat{L}_j\hat{L}_k\rangle\langle\hat{L}_i\rangle +
\langle\hat{L}_k\hat{L}_i\rangle\langle\hat{L}_j\rangle \nonumber \\
& - & 2\langle\hat{L}_i\rangle\langle\hat{L}_j\rangle\langle\hat{L}_k\rangle .
\end{eqnarray}

The equations of motion for the expectation values $l_i\equiv \langle \hat{L}_i
\rangle$ and the second order moments $\Delta_{ij}\equiv \langle 
\hat{L}_i\hat{L}_j + \hat{L}_j\hat{L}_i \rangle - 2\langle\hat{L}_i\rangle
\langle\hat{L}_j\rangle$ 
are then obtained from (\ref{fullqmeqns}). We are interested in the case
when all atoms start in mode $1$, i.e.
$l_0(0)=N,l_x(0)=l_y(0)=0,l_z(0)=-N/2$ and
$\Delta_{xx}(0)=\Delta_{yy}(0)=N/2$ (all other second moments vanish) and
for simplicity we take $\gamma_1=\gamma_2=\Gamma$, $\gamma=0$.
Writing only the relevant equations and omitting vanishing terms (such as those
proportional to $\Delta_{xz}$ and $\Delta_{yz}$ which are zero for all times),
we have (after some algebra)

\begin{eqnarray}
\dot{l}_0 &=& -2\Gamma l_0 \nonumber \\
\dot{l}_z &=& -2\Gamma l_z -\chi(\Delta_{xx}-\Delta_{yy}) \nonumber \\
\dot{\Delta}_{xx} &=& -4\Gamma\Delta_{xx}+\Gamma l_0 +4\chi l_z\Delta_{xx}
\nonumber \\
\dot{\Delta}_{yy} &=& -4\Gamma\Delta_{yy}+\Gamma l_0 -4\chi l_z\Delta_{yy}
\label{bbgky1}
\end{eqnarray}

and $l_x(t)=l_y(t)=0$. These equations are non-linear and cannot be solved
analytically nor perturbatively in $\Gamma/\chi$. For short enough times,
the number of excitations into mode $2$ is small and $l_z\simeq -N/2$, so
that plugging this in (\ref{bbgky1}) we have

\begin{eqnarray}
\Delta_{xx}(t) &\simeq& {\tiny \frac{N}{2}}e^{-2N\chi t}+O[\Gamma/\chi]
\nonumber \\
\Delta_{yy}(t) &\simeq& {\tiny \frac{N}{2}}e^{2N\chi t}+O[\Gamma/\chi]
\end{eqnarray}

i.e. the variance of the x-component of the pseudo-angular momentum is
squeezed while
that of the y-component is anti-squeezed. Plugging these back in the equation 
of motion of $l_z$, we obtain $l_z(t)\simeq -N/2+(\cosh{2N\chi t}-1)/2+
O[\Gamma/\chi]$. This equation predicts growth of $l_z$ without bound, however
we know that because $l_z$ is the z-component of an angular momentum vector, 
we must have $|l_z|\leq N/2$. The phase space of this angular momentum vector 
is the Bloch sphere and in essence we have neglected the small
curvature of the Bloch sphere (of radius $R=N/2$) and have approximated
the 
phase space 
by the flat planar phase space of a harmonic oscillator. We call this 
approximation the bosonic approximation, since it predicts infinite
squeezing in the long time limit and in the absence of dissipation,
similar to the case of squeezed light. 
Formally this is equivalent to assuming

\begin{eqnarray}
[\hat{a}_2^\dagger \hat{a}_1,\hat{a}_1^\dagger\hat{a}_2] &=& 
\hat{a}_2^\dagger\hat{a}_2-\hat{a}_1^\dagger\hat{a}_1 \nonumber \\
&\simeq& -N
\end{eqnarray}

i.e. the operator $\hat{S}_+=\hat{a}_2^\dagger \hat{a}_1/\sqrt{N}$ obeys 
bosonic commutation relations. Under this approximation the Hamiltonian 
(\ref{ctham}) becomes $\hat{H}=-i(\hbar\chi N/2)(\hat{S}_+^2-\hat{S}_-^2)$
which
is identical to the Hamiltonian describing squeezing of light 
\cite{Walls-Milburn}.

In order to take into account the curvature of phase space and the non-bosonic
nature of the angular momentum operators, we use the following
transformation

\begin{eqnarray}
\hat{N} &=& N \hat{h}_0 \nonumber \\
\hat{L}_x &=& \sqrt{N} \hat{h}_x \nonumber \\
\hat{L}_y &=& \sqrt{N} \hat{h}_y \nonumber \\
\hat{L}_z &=& \hat{h}_z-\frac{N}{2}\hat{h}_0 
\label{groupcon}
\end{eqnarray}

in terms of which the commutation relations become

\begin{eqnarray}
\left[ \hat{h}_{x,y,z},\hat{h}_0 \right] & = & 0 \nonumber \\
\left[ \hat{h}_z,\hat{h}_{\pm} \right] & = & \pm \hat{h}_{\pm} \nonumber \\
\left[ \hat{h}_+,\hat{h}_- \right] & = & 2\frac{\hat{h}_z}{N}-\hat{h}_0
\end{eqnarray}

where $\hat{h}_\pm=\hat{h}_x\pm i\hat{h}_y$. In the limit $N\rightarrow \infty$
these commutation relations become those of bosonic operators i.e. 
$\lim_{N\rightarrow\infty}[\hat{h}_0,\hat{h}_z,\hat{h}_+,\hat{h}_-]=[\hat{1},
\hat{a}^\dagger\hat{a},\hat{a}^\dagger,\hat{a}]$, a process formally known as a
group contraction \cite{Courtens}. The linear transformation of operators
(\ref{groupcon}) does not introduce any extra approximation.
 
The Hamiltonian (\ref{ctham}) can be re-expressed as 

\begin{eqnarray}
\hat{H} &=& \hbar\chi N\left(\hat{h}_x\hat{h}_y+\hat{h}_y\hat{h}_x\right)
\nonumber \\
&=& -i\frac{\hbar\xi}{2}\left(\hat{h}_+^2-\hat{h}_-^2\right)
\end{eqnarray}

where we have defined $\xi=\chi N$.
We can now obtain equations of motion for the expectation
values $h_j=\langle \hat{h}_j \rangle$ and the second order moments 
$\delta_{ij}=\langle \hat{h}_i\hat{h}_j \rangle -2 \langle \hat{h}_i \rangle
\langle \hat{h}_j \rangle$ from (\ref{bbgky1}). Letting $\tau=N\chi
t=\xi t$ be a rescaled time, $\kappa=\Gamma/(N\chi)=\Gamma/\xi$ be
the rescaled dissipation rate and writing $\epsilon=1/N$ and $\dot{x}={\rm
d}x/{\rm d\tau}$, 
we have 

\begin{eqnarray}
\dot{h}_0 &=& -2\kappa h_0 \nonumber \\
\dot{h}_z &=& -2\kappa h_z -(\delta_{xx}-\delta_{yy}) \nonumber \\
\dot{\delta}_{xx} &=& -4\kappa\delta_{xx}+\kappa h_0-2h_0\delta_{xx}+4\epsilon 
h_z\delta_{xx} \nonumber \\
\dot{\delta}_{yy} &=& -4\kappa\delta_{yy}+\kappa h_0+2h_0\delta_{yy}-4\epsilon 
h_z\delta_{yy}. 
\label{bbgky2}
\end{eqnarray}

Note that these equations are formally equivalent to (\ref{bbgky1}), no 
approximation has been made from (\ref{bbgky1}) to (\ref{bbgky2}). Letting
$\epsilon\rightarrow 0$ reproduces the results of the bosonic
approximation obtained
above in the limit of $l_z\simeq -N/2$. Terms of order $\epsilon$ and higher
represent corrections to the bosonic approximation and, as shown below,
they give rise to a limit to the amount of squeezing achievable. 

Solving (\ref{bbgky2}) to first order in $\epsilon$ and $\kappa$ we
obtain, writing only the relevant terms,

\begin{eqnarray}
\delta_{xx}(\tau) &=& \frac{1}{2}\left[e^{-2\tau}+(\kappa+\epsilon/2)+\cdots
\right]
\label{squbbgky}
\end{eqnarray}

which shows that the variance $\Delta J_x=\sqrt{(N/2)\delta_{xx}}$ is squeezed.
 Second order terms in $\kappa$ and $\epsilon$ come multiplied by an
exponentially 
growing term $e^{2\tau}$ so that as a function of time, the variance reaches 
a minimum value $\delta_{xx}\sim {\rm max}[\kappa,\epsilon]$ at a time 
$e^{-\tau_*}\sim {\rm max}[\kappa,\epsilon]$, after which it grows 
exponentially and the squeezing is lost. Note that this behaviour
($\delta_{xx}(t)$ reaches a minimum value and then increases again) also
occurs when $\kappa\rightarrow 0$, indicating that it is a generic feature
of the finite system size.
This model predicts that
a variance $\delta_{xx}\sim \epsilon=1/N \rightarrow \Delta J_x^2\sim 1$
is achievable as long as losses are small enough, i.e.
$\kappa\sim\epsilon$, 
which in terms of $\chi$ and $\Gamma$ means

\begin{equation}
\chi \sim \Gamma\;\;\;{\rm or}\;\;\xi \sim N\Gamma
\label{heislimit}
\end{equation}

where $\chi$ corresponds to the single-atom nonlinear interaction rate and
$\Gamma$ represents the single-atom loss rate.

In order to achieve any squeezing ($\delta_{xx}\leq 1/2$) it is
necessary to have $\kappa<1$ i.e. $N\chi > \Gamma$. 
In the regime $N\chi \gg \Gamma$ very
strong correlations can be obtained. Note that the single-atom
nonlinearity can still be relatively weak compared to the single particle
loss rate ($\chi\ll\Gamma$). 
For example when the dissipation rate is such that
$\kappa\sim\sqrt{\epsilon}$ i.e. $\sqrt{N}\chi\sim\Gamma$, the amount of
squeezing obtained (\ref{squbbgky}) is $\delta_{xx}\sim 1/\sqrt{N}$. It
takes a time $e^{-2\tau}\sim\sqrt{\epsilon}$ to reach this state and the
number of particles lost during that time is $\Delta N\sim N\times
2\kappa\tau\sim\sqrt{N}\log{N}$. 
This number can therefore also be thought of
as the maximum number of particles that can be lost from the ensemble
without destroying squeezing beyond $\delta_{xx}\sim 1/\sqrt{N}$.

In order to reach the Heisenberg limit it is required that the
single atom nonlinearity $\chi$ be larger than the decay rate $\Gamma$. 
Note that in this case, the number of atoms lost by the time optimal
squeezing is achieved is $\Delta N\sim\log{N}$
which indicates that a very small number of atoms is lost. This
number also corresponds to the maximum number of particles that can be
lost from the ensemble and not destroy squeezing at the Heisenberg limit
level. Clearly the more squeezed the state of the atoms is, the more
sensitive it becomes to atom loss and in general to any form of
dissipation.

\section{Coherent atom interactions via slow light}

We now describe a technique to induce effective coherent 
interactions between atoms in metastable states \cite{Paper1}. 
The technique is based on a resonantly enhanced
nonlinear process involving Raman scattering into a 
``slow'' optical mode \cite{slow-light}, which creates a 
pair of spin-flipped atom and slowly propagating coupled excitation
of light and matter (dark-state  polariton). When the group
velocity of the polariton is reduced to zero \cite{stop-light,darkpolar}, 
this results in pairs of spin flipped atoms. 
The fact that pairs of atomic excitations are created in this
process can also be viewed as a coherent interaction between atoms, i.e. a 
controlled ``collision'' leading to entanglement of the state of each atom 
with that of every other atom in the ensemble. 

A number of proposals 
have been made for generating entangled states of atomic ensembles 
and resulting in so-called spin squeezed states.  Some are based 
on interatomic interactions at ultra-cold temperatures \cite{squ-cold}, whereas
others involve  mapping the states of non-classical light fields into 
atoms \cite{squ-Polzik}, QND measurements of spins \cite{squ-Bigelow} with 
light or dipole blockade for Rydberg atoms \cite{Bouchoule}. 
In contrast to some of these mechanisms the present approach does not require  
coherence of the atomic motion or sources of non-classical light 
and  is completely deterministic thereby 
significantly simplifying possible experimental realizations. 
We further show that the present technique can be made robust with 
respect to realistic decoherence processes such as spontaneous emission 
and leakage of slow photons from the medium.

We consider a system of $N$ atoms (Fig. 4a) interacting with two 
classical driving fields $\Omega_{1,2}$ and
one quantized mode $\hat{a}$ of a running wave cavity that is initially in a 
vacuum state. Note that we consider a cavity configuration for ease of 
theorerical treatment, the results of this analysis however remain valid in the
limit of unity finesse, i.e. in free space configuration.
Relevant atomic sublevels
include two manifolds of metastable states (e.g hyperfine sublevels of 
electronic ground state) and excited states that may be accessed by optical 
transitions. The atoms are initially prepared in their ground states 
$|g\rangle$. One of the classical fields, of Rabi frequency $\Omega_1$, 
is detuned from the atomic resonance by an amount roughly equal to the 
frequency splitting between ground state manifolds. 
The other field of Rabi frequency $\Omega_2$ is resonant with an atomic 
transition $|b_2\rangle\rightarrow|a_2\rangle$. 
The quantized field can be involved in two Raman transitions 
corresponding to Stokes and anti-Stokes processes. Whereas the former
corresponds to the usual Stokes scattering in the forward direction, the 
latter establishes an Electromagnetically Induced Transparency (EIT) and 
its group velocity is therefore substantially reduced.


\begin{figure}[ht]
\begin{center}
\epsfig{file=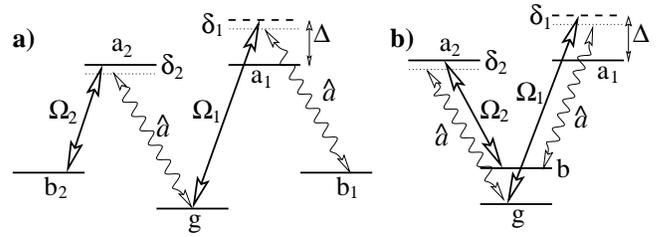,width=8.5cm}
\leavevmode
\end{center}
\vspace*{2ex}
\caption{
Energy levels scheme for the effective coherent interaction leading to creation
of pairs of atoms a) in different final states (``non-degenerate'' scheme) and
b) in identical final states(``degenerate'' version).}
\end{figure}


The pair excitation can be viewed as resulting from quantized photon 
exchange between atoms (Fig. 5) in a two-step process.  The first flipped
spin is created due to  Stokes Raman scattering, which also results in 
photon emission in a corresponding Stokes mode. In the presence of EIT,
this photon is directly converted into a dark-state polariton which becomes 
purely atomic  when the group velocity is reduced to zero.  
This implies that atomic spins are always flipped in pairs. 
In Fig. 4a the two final states involved in Raman 
transitions are different and atomic pairs in different states
are created. In Fig. 4b the final states of the two Raman processes are 
identical, in which case  atomic pairs in the same state result. The analysis
of this ``degenerate'' version of the scheme is similar to the 
``non-degenerate'' case and we will consider only the latter case 
here.


\begin{figure}[ht]
\begin{center}
\epsfig{file=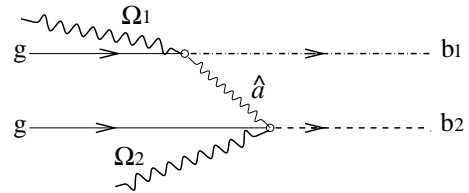,width=6.0cm}
\leavevmode
\end{center}
 \vspace*{2ex}
 \caption{Diagram illustrating coherent atom-atom interaction mediated by
dark-state polariton, leading to the creation of a pair of spin-flipped atoms.}
\end{figure}


For conceptual simplicity we assume that the quantized field 
corresponds to  a single mode  of a running-wave cavity with a creation 
operator $\hat{a}^\dagger$ and atom-field coupling constants $g_1$ and
$g_2$.  
The interaction
Hamiltonian for the system of N atoms and light can be split into two 
parts $H = H_{Ram} + H_{res}$ corresponding to the Stokes Raman process and the
anti-Stokes process respectively:
 
\begin{eqnarray}
H_{Ram}&=&-\hbar\Delta \Sigma_{a1a1} - \hbar\delta_1 \Sigma_{b_1b_1} 
\nonumber \\
&+&[\hbar\Omega_1 \Sigma_{a_1g} 
+ \hbar g_1 a \Sigma_{a_1b_1} + {\rm
h.c.}], \label{hamram} \\
H_{res}&=&\hbar\delta_2 \Sigma_{b_2b_2} + \hbar\delta_2 \Sigma_{a_2a_2} 
\nonumber \\
&+& [\hbar g_2 a \Sigma_{a_2g} + \hbar\Omega_2 \Sigma_{a_2b_2} + {\rm h.c.}], \label{hamres}
\end{eqnarray}

where $\Sigma_{\mu\nu} = \sum_i 
|\mu\rangle_{ii}\langle \nu|$ are collective atomic operators corresponding to 
transitions between atomic states $|\mu\rangle,|\nu\rangle$, $\Delta$ is the
detuning of the classical field $\Omega_1$ from the single-photon transition 
$|g\rangle \rightarrow |a_1\rangle$, $\delta_1$ and $\delta_2$ are the 
two-photon detunings from the $|g\rangle \rightarrow |b_1\rangle$ and
$|g\rangle \rightarrow |b_2\rangle$ transitions respectively as shown in
Fig. 4.

In the limit of large detuning $\Delta$ and ignoring two-photon detunings for 
the moment, the Hamiltonian $H_{Ram}$ describes
off-resonant Raman scattering. 
We take into account realistic decoherence
mechanisms such as spontaneous emission from the 
excited states in all directions and decay of the cavity mode 
with a rate $\kappa$. The evolution of atomic operators is then described 
by Heisenberg-Langevin equations:
\begin{eqnarray}
{\dot \Sigma}_{\mu\nu} = -\gamma_{\mu\nu} \Sigma_{\mu\nu} + {i \over \hbar}
[H, \Sigma_{\mu\nu}] + F_{\mu\nu}, 
\end{eqnarray}
where $\gamma_{\mu\nu}$ is a  decay rate of coherence $\mu\rightarrow \nu$
and $F_{\mu\nu}$ are associated noise forces. The latter have zero average 
and are $\delta$-correlated with associated diffusion coefficients that 
can be found using the Einstein relations. 

After a canonical transformation corresponding to adiabatic  elimination of
the excited state (see Appendix C for details), $H_{Ram}$ becomes equivalent
to the effective Hamiltonian

\begin{equation}
{\tilde H}_{Ram} = \hbar\chi\hat{a}^\dagger\hat{S}_1^\dagger
+ {\rm h.c.}
\label{hamrameff}
\end{equation}

where $\hat{S}_1=\Sigma_{gb_1}/\sqrt{N}$ and 
$\chi=g_1\sqrt{N}\Omega_1^*/\Delta$.
This effective hamiltonian thus describes the process in which a Stokes
photon is emitted necessarily accompanied by a spin flip. The quantum
state of the Stokes mode is thus perfectly correlated with the state of
the atomic spin flip mode.

The resonant part of 
the Hamiltonian $H_{res}$ is best analyzed in terms of dark and bright-state 
polaritons \cite{brightpolar}
\begin{eqnarray}
P_D &=& {\Omega_2 a - g_2 \sqrt{N} S_2 \over \sqrt {g_2^2 N + \Omega_2^2}}, \nonumber \\
P_B &=& {g_2 \sqrt{N} a + \Omega_2 S_2 \over \sqrt {g_2^2 N + \Omega_2^2}},  
\label{polaritons}
\end{eqnarray}
which are superpositions of photonic and atomic excitations $\hat{a}$ and 
$S_2=\Sigma_{gb_2}/\sqrt{N}$.. 
In particular, $H_{res}$
has an important family of dark-states: 
\begin{eqnarray}
|D^n\rangle \sim (P_D^\dagger)^n |g\rangle|{\rm vac}\rangle
\label{dark}
\end{eqnarray}
with zero eigenenergies. This means that once in the dark state, the system 
stays in the dark state. Note that all other eigenstates of $H_{res}$ 
have, in general, non-vanishing interaction energy.
Under conditions of Raman resonance and sufficiently slow excitation 
(``adiabatic condition'', see Appendix D for details) 
the Stokes photons emitted by Raman scattering, Eq.(\ref{hamrameff}),  will 
therefore couple  solely to the dark-states (\ref{dark}). In this case
the coherent part of the evolution of the entire system is described by an
effective 
Hamiltonian: 
\begin{eqnarray}
H_{eff} = -i \hbar \xi (P_D^\dagger S_1^\dagger - S_1 P_D), 
\label{eff} 
\end{eqnarray}
with $\xi = \Omega_1 \Omega_2/ \Delta  \times g_1 \sqrt{N}/  
\sqrt{g_2^2 N + \Omega_2^2}$ (without loss of generality, $\xi$ was chosen
imaginary here for simplified calculations). 
The Hamiltonian (\ref{eff}) describes the coherent
process of generation of pairs of excitations involving polaritons and 
spin-flipped atoms. 
Note that for small number of excitations the spin waves and
polaritons obey bosonic commutation relations and this Hamiltonian is formally 
equivalent to that describing optical parametric amplification (OPA) of two
modes \cite{Walls-Milburn}. 
In the non-bosonic limit, this Hamiltonian is also analogous to the 
``counter-twisting'' model of (\ref{ctham}).
In appendix D we show that the coupled equations for the polariton $P_D$
and the spin flip $S_1$ are given by 

\begin{eqnarray}
\dot{S}_1^\dagger &=& (\frac{|g_1|^2}{|g_2|^2}\gamma_L-\gamma_L-i\delta_1)
S_1^\dagger + \xi P_D +{\tilde F}_{S_1}^\dagger(t) 
\label{finspin}
\\
\dot{P}_D &=& -(\kappa/\eta+\gamma_L+i\delta_2)P_D +\xi S_1^\dagger +
{\tilde F}_D(t)
\label{finpol}
\end{eqnarray}

where the polariton decay rate includes an atomic part $\gamma_L$ and a
photonic part $\kappa/\eta$ due to leakage of photons out of the medium
(at a rate reduced by the factor $\eta=|g_2|^2N/|\Omega_2|^2$ equal to  
the ratio of vacuum light velocity to the group
velocity of slowly propagating Stokes photons). 
The spin flip operator
equation (\ref{finspin}) is seen to
contain both a decay term and a gain term due to spontaneous emission into 
the bright polariton mode. Note that this apparent decrease in dissipation
is however accompanied by increased fluctuations denoted by the new noise
force operator ${\tilde F}_{S_1}(t)$.
The effective detuning between
the polariton and spin flip mode is seen to correspond to the difference
in two-photon detunings
$\delta_1-\delta_2$.

We now consider the scenario in which the system is evolving for a time $\tau$ 
under the Hamiltonian $H_{eff}$, after which both fields are turned off. 
If the  procedure is adiabatic  upon turn-off of the coupling fields 
$\Omega_{1,2}$ the polaritons are converted into pure spin excitations 
$P_D \rightarrow S_2$. Hence the entire procedure will correspond to the 
following state of the system:
\begin{eqnarray}
|\Psi\rangle = && { 1\over \cosh\xi\tau} \sum_n (\tanh\xi\tau)^n {1 \over n!}
(P_D^\dagger)^n (S_1^\dagger)^n |g\rangle|{\rm vac}\rangle \nonumber \\
 && \rightarrow 
{ 1\over \cosh\xi\tau} \sum_n (\tanh\xi\tau)^n |n_{b_1},n_{b_2} \rangle|{\rm vac}\rangle.
\end{eqnarray}
Here $|n_{b_1},n_{b_2} \rangle = 1/n! (S_2^+)^n (S_1^+)^n |g\rangle$ are 
Dicke-like 
symmetric states of atomic ensemble and we assumed $n_{b_1,b_2} \ll N$.
For non-zero $\xi \tau$ this state describes an entangled state, 
for which relative fluctuations between the two modes 
decreases exponentially to values well below the standard quantum limit 
(SQL) corresponding to uncorrelated atoms.

The present technique can also be viewed as a new mechanism for 
coherent ``collisions'' \cite{Meystre}
between atoms mediated by light. In particular, the case when 
atomic pairs are excited into two different levels 
(as e.g. in Fig. 4a) closely resembles coherent spin-changing 
interactions that occur in degenerate atomic samples \cite{Ketterle}, 
whereas the case
when atomic pairs are stimulated into the identical state (Fig. 4b) 
is reminiscent of dissociation of a molecular condensate 
\cite{Heinzen}. 
To put this analogy in perspective we note 
that the rate of the present optically induced process can exceed that of 
weak interatomic interactions by orders of magnitude. Therefore the 
present work may open up interesting new possibilities 
for studying  many-body phenomena of strongly interacting atoms. 


\begin{figure}[ht]
\centerline{\epsfig{file=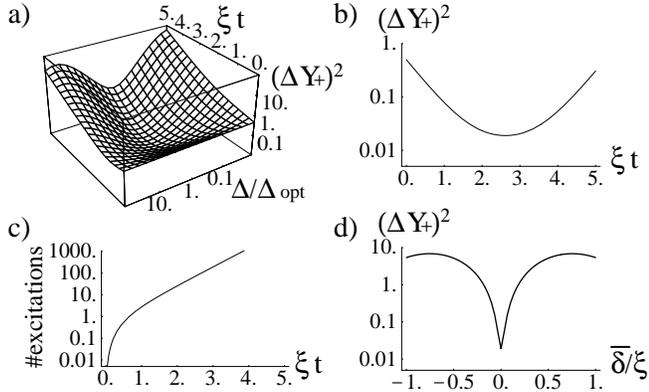,width=8.5cm}}
 \vspace*{2ex}
 \caption{(a) Quadrature variance $\Delta Y_+^2$ vs. single-photon detuning
$\Delta$ and interaction time $\xi t$, (b) same for $\Delta=\Delta_{opt}$ and
$\delta_1=\delta_2$ showing maximum squeezing $\Delta Y_+^2\simeq 0.02$
(for $\sqrt{g^2N/\gamma\kappa}=100$), (c) Number of excitations pumped
in the system vs. time (same conditions as in b) and (d) $\Delta Y_+(t^*)^2$
vs. two-photon detuning ${\bar \delta}\equiv (\delta_1-\delta_2)/2$ for
$\Delta=\Delta_{opt}$ and where $t^*$ gives maximum squeezing.}
\end{figure}


To quantify the resulting correlations established between the polariton mode
$P_D$ and the pure spin flip mode $S_1$, we introduce the quadratures of both
modes (which are bosonic for small number of excitations) in direct analogy to
the optical parametric case. We define 
the quadratures $X_1=(S_1+S_1^+)/\sqrt{2}$, $Y_1=i(S_1-S_1^+)/\sqrt{2}$
and similarly for the polarition mode;
these can be measured e.g. by converting spin excitations to light. 
Correlations between the modes appear due to dynamical evolution and 
squeezing is found in certain quadratures of the sum and difference modes 
$X_-=(X_1-X_D)/\sqrt{2}$ and $Y_+=(Y_1+Y_D)\sqrt{2}$. In the language of
harmonic oscillators, the positions in mode $1$ and $2=D$ are correlated 
($X_1\simeq X_D$), while the momenta are anti-correlated ($Y_1\simeq -Y_D$).
For small number 
of excitations the sum and difference modes obey standard commutation 
relations $\left[X_\alpha,Y_\beta\right]=-i\delta_{\alpha,\beta}$ 
where $\alpha,\beta=+,-\,{\rm or}\,1,D$. A quadrature $Y_\pm$ is squeezed 
when  $\Delta Y_\pm(t)^2<1/2$ and the Heisenberg limit corresponds to
$\Delta Y_\pm(t)^2\sim 1/N$.  

We find that squeezing is optimal under 
conditions of four-photon resonance 
($\delta_1=\delta_2$) and in the limit of $\eta\gg 1$ (Fig. 6).   
Evolution leads to squeezing of $Y_+$ and $X_-$, anti-squeezing of $Y_-$
and 
$X_+$. The squeezing in $Y_+$ reaches a minimum value at $t=t^*$ after 
which the growing fluctuations in $X_+$ give rise to increased noise in 
$Y_+$. 
Note that the total number of excitations (both modes) in the system, 
equal to $\langle X_+^2+X_-^2+Y_+^2+Y_-^2 \rangle$,
grows exponentially with time (Fig. 6c). 
Specifically, in the case $g_1=g_2=g$ and thus
$\gamma_1=\gamma_2\equiv\gamma$, for $\xi t>1$, we have:

\begin{eqnarray}
\left(\Delta Y_+(t)\right)^2 &=&1/2\left\{e^{-2\xi t} + 
\frac{3\gamma_L+\kappa/\eta}{2\xi}\right. \\ \nonumber
&+& \left. \left(\frac{\gamma_L+\kappa/\eta}{2\xi}\right)^2e^{2\xi t}\right\}
\label{quadt}
\end{eqnarray}

where we have neglected terms of higher order in $\gamma_L/\xi$ and 
$\kappa/(\eta\xi)$.
The maximum amount of squeezing is obtained after an interaction time 
$t^*$ such that $e^{-2\xi t^*}=(\gamma_L+\kappa/\eta)/2\xi$ and is given by 

\begin{equation}
(\Delta Y_+)^2=\frac{5\gamma_L+3\kappa/\eta}{4\xi}
\label{squeez1}
\end{equation}

i.e. of the order of the damping rate divided by the coherent interaction rate.

Since both the interaction parameter $\xi$ and the relaxation rate of the
polariton $\gamma_D=\gamma_L+\kappa/\eta$ depend on the single photon
detuning $\Delta$ (Fig. 6a), we find that squeezing is optimized for 
\begin{equation}
\Delta_{opt}=\gamma\sqrt{\frac{5|\Omega_1|^2}{3|\Omega_2|^2}\frac{|g|^2N}
{\gamma\kappa}}
\end{equation}
and with this optimal value of the detuning, the squeezing reaches a minimum
value of  
\begin{equation}
\left(\Delta Y_{+opt}\right)^2 = \frac{\sqrt{15/4}}
{\sqrt{|g|^2N/\gamma\kappa}}.
\label{maxsqueez}
\end{equation} 
Note that the denominator is equal to the atomic
density-length product multiplied by the empty cavity finesse and can
easily 
exceed $10^4$ even for modest values of the density-length product and
cavity finesse. We further emphasize that typical generation rate 
resulting in such optimal squeezing
$\Omega_1\Omega_2/\Delta_{opt}$ can easily be on the order of fraction of
MHz.  In such a case other decoherence mechanisms are negligible.    
Doppler shifts can also be disregarded as long as all fields are 
co-propagating. 

For the ``degenerate'' version of the interaction (i.e. with identical final 
states for the spin flips, see Fig. 4b), the effective hamiltonian can be
written as

\begin{equation} 
H_{eff} = i\hbar\xi(\hat{S}^{\dagger 2}-\hat{S}^2)
\end{equation}

where the limit $\eta\gg 1$ has been used to write $P_D\simeq -S$, with
$S=1/\sqrt{N}\Sigma_{gb}$ the spin flip operator. In this 
case the correlations lead to squeezing of $X=(S+S^\dagger)/\sqrt{2}$ and
anti-squeezing of $Y=i(S-S^\dagger)/\sqrt{2}$. The analysis for this
configuration is
very similar to the non-degenerate version, in particular the maximum amount
of squeezing achievable is also given by an expression of the form 
(\ref{maxsqueez}).

We can now obtain a condition for achieving Heisenberg-limited spin
squeezed
states, i.e. $\left(\Delta Y_{+opt}\right)^2\simeq 1/N$. We see from 
(\ref{squeez1}) that this requires

\begin{equation}
\xi \sim N\Gamma
\end{equation}

where $\Gamma=(5\gamma_L+4\kappa/\eta)/4$ is the effective damping rate of the
system. This is in complete agreement with the estimate based on our
simple bosonic model of section III (\ref{heislimit}). 
In terms of the single photon Rabi
frequency $g$, the cavity decay rate $\kappa$, the spontaneous emission rate 
$\gamma$ and the number of atoms $N$, the condition for achieving some
squeezing i.e. $(\Delta Y_+)^2<1/2$ is

\begin{equation}
|g|^2N > \kappa\gamma
\end{equation}

which can be easily achieved in the laboratory since it simply
corresponds to the condition that the
density length product multiplied by the cavity finesse be larger than 
one. 
In the cavity QED regime of strong coupling $|g|^2\sim\kappa\gamma$, very
strong quantum correlations i.e. $(\Delta Y_+)^2\sim 1/\sqrt{N}$
between atoms can be produced. 
In order to obtain Heisenberg limited spin squeezed states i.e.
$(\Delta Y_+)^2 \sim 1/N$, one requires a more stringent condition

\begin{equation}
|g|^2\sim N\kappa\gamma
\end{equation}

which can be fulfilled only in the strong coupling regime of cavity QED
for a limited number of atoms. Note that this regime has been achieved
experimentally by
several groups \cite{cavityQED} and would allow for Heisenberg limited
spin squeezing for as many as $\sim 10^3$ atoms.
We have analyzed in this paper the situation of a running-wave
cavity, so that all atoms couple equally apart from a possible phase to
the cavity mode irrespective of their position.
In order to fulfill the cavity QED regime, small cavity volume is needed
i.e. standing wave cavities.
For atoms in such a cavity the coupling to the cavity mode is position
dependent and it becomes necessary to localize atoms accurately at the 
antinodes of a trapping mode. Note that significant experimental progress
has been made towards this direction by several groups \cite{1atom}. 
Once the atoms are well localized in the cavity,
the interaction can proceed via a neighbouring mode $b$ (e.g. different
from the trapping mode $a$) so that
for atoms localized within a small region in the cavity the two modes have
essentially the same wavelength and atoms would therefore couple equally
to the $b$ mode as well, irrespective of their position. 

\section{Discussion and Conclusion}

We have reviewed Ramsey spectroscopy and the use of spin squeezed states
in precision measurements of this type. With the experimental motivation
of minimizing the phase accuracy in phase estimation with Ramsey
fringes, we introduced a particular class of squeezed states. These states
lead to Heisenberg limited phase accuracy and we developed various
pictorial representations for them. The strong similarities of these
representations of spin squeezed states to those of squeezed states of
light suggests an analogy extending to the type of interaction that gives
rise to squeezing. We are thus lead to consider the so-called "counter
twisting" Hamiltonian, which has been shown to lead to maximal spin
squeezing. We have studied this model for spin squeezing in the presence
of a dissipation mechanism and analyzed the effect of damping and finite
system size on the amount of squeezing achievable with such an
interaction. The analysis was based on a decorrelation approximation to
the BBGKY hierarchy of equations of motion, followed by the use of a
linear transformation which in the limit of large number of atoms
$1/N\rightarrow 0$ "contracts" the angular momentum operators onto bosonic
operators. This allows for the systematic inclusion of finite system
size effects. It appears that Heisenberg limited spin squeezed states may
be produced when the single atom nonlinearity exceeds the single atom loss
rate. In this case the maximum number of atoms that can be lost before
quantum correlations are destroyed to the point of compromising the spin
squeezing is of the order $\Delta N\sim\log{N}$. For spin squeezing at a
more modest level than the Heisenberg limit, larger number of atoms may be
lost without compromising the squeezing, indicating the stronger 
sensitivity of spin squeezed states to dissipation for larger amounts of
squeezing. 

We have also presented in detail a scheme based on the interaction of
coherent classical light with an optically dense ensemble of atoms that
leads to an effective coherent spin-changing interaction involving pairs
of atoms. Atoms may be transferred to the same final state leading to spin
squeezing (analogous to squeezing of light by degenerate OPO) or to
different final states in this case leading to quantum correlations
between
different atomic modes (analogous to quantum correlations between
electromagnetic modes by non-degenerate OPO). 
We have shown that this process is robust with respect to realistic 
decoherence mechanisms and can result in rapid generation of correlated 
(spin squeezed) atomic ensembles. 
The amount of correlations created by this effective interaction can be
simply expressed in terms of the single photon Rabi frequency $g$, the
atomic spontaneous emission rate $\gamma$ and the cavity decay rate
$\kappa$. 
We find that
the generation of spin squeezed states requires $g^2N\sim\kappa\gamma$,
which can easily be achieved in low finesse cavities with e.g. room
temperature atomic vapours. Very strongly correlated states can be
produced when the strong coupling regime $g^2\sim\kappa\gamma$ of cavity
QED is achieved and the generation of Heisenberg limited spin squeezed
states requires $g^2\sim N\kappa\gamma$.
The effective interaction rate $\xi=\Omega_1\Omega_2/\Delta$ which depends
on the Rabi-frequency of two applied classical fields $\Omega_{1,2}$ and a
detuning from an atomic transition $\Delta$ can be fast and is
controllable.
Furthermore, the resulting spin excitations can be easily converted into 
photons on demand, which facilitates applications in quantum information
processing. 
Possible applications involving high-precision measurements in atomic clocks 
can be also foreseen.

We thank M.Fleischhauer, J.I.Cirac, V.Vuletic, S.Yelin and P.Zoller 
for helpful  discussions. 
This work was supported by the NSF through the grant to the ITAMP. 


\appendix

\section{Ramsey Spectroscopy}

In Ramsey spectroscopy \cite{Wineland}, a collection of $N$ two-level atoms 
are made to interact 
with two separated fields (in time or in space). The lower and upper 
states (refered to as ground and excited state) have
an energy difference $\hbar\omega_0$ and atoms will thus acquire a different 
dynamical phase $e^{-i E t/\hbar}$ depending on which state they are in. 
The effect of properly chosen electromagnetic fields is to perform a 
transformation that prepares the atoms 
in a superposition of the two states $|g\rangle$ and $|e\rangle$. 
The different
parts of the wavefunction of atoms (corresponding to the ground and excited
state) acquire a relative phase due to dynamical evolution and when the 
inverse transformation is applied, an interference effect is obtained.  
An exact parallel with the Mach-Zender interferometer can be drawn
\cite{Dowling}: 
the transformation preparing atoms in a superposition of ground and excited 
states
is equivalent to the transformation that lets a photon incident 
on a beam splitter
explore the two arms of an interferometer. The relative phase acquired in
the two atomic states during free evolution of duration $T$ is the equivalent 
of the relative phase acquired by photons travelling in the arms of the 
interferometer.
Finally, the second pulse that performs the inverse
transformation on atoms is the equivalent of the recombination of signals from 
the two interferometer arms on a beam splitter. 
At the end of this sequence, the
number of atoms in either states, equivalent to the number of photons
from either output of the final beam splitter, is measured. 
In this way, the signal 
measured depends on the acquired relative phase which can thus be estimated 
with some accuracy.

We will now quantify this more precisely: let the frequency of the applied 
electromagnetic pulses be $\omega$, 
and the time delay between the two zones of interaction be $T$. 
The duration and strength of the applied fields are
chosen so as to lead to $\pi/2$ pulses, i.e. transformation of the
atomic state according to

\begin{eqnarray}
|e\rangle & \rightarrow & \frac{|e\rangle -i |g\rangle}{\sqrt{2}}
\nonumber \\
|g\rangle & \rightarrow & \frac{|e\rangle +i |g\rangle}{\sqrt{2}}.
\label{pi2pulse}
\end{eqnarray}

During their free evolution between the two zones, atoms in the ground and
excited states acquire a relative phase $\phi$ which, in a frame rotating
with the frequency of the applied field, is $\phi=(\omega-\omega_0)T$.

Before entering the first interaction zone, the atoms are prepared in
their lower state $|g\rangle$ and at the exit of the second zone, the
number of atoms in states $|e\rangle$ and $|g\rangle$ is measured. 

For simplicity, we consider the case when the first zone leads to a $\pi/2$
pulse and the second one a $-\pi/2$ pulse. 
The picture of angular momentum is particularly well suited to discuss the
Ramsey interferometric scheme and leads to an intuitive pictorial 
representation of the scheme.
The Schwinger angular momentum operators are defined as

\begin{eqnarray}
\hat{J}_x & = & (\hat{\Sigma}_{eg}+\hat{\Sigma}_{ge})/2 \nonumber \\
\hat{J}_y & = & (\hat{\Sigma}_{eg}-\hat{\Sigma}_{ge})/2i \nonumber \\
\hat{J}_z & = & (\hat{\Sigma}_{ee}-\hat{\Sigma}_{gg})/2
\end{eqnarray}

where $\hat{\Sigma}_{\mu\nu}=\sum_{j=1}^{N}|\mu\rangle_{jj}\langle \nu|$
are collective operators. In terms of these, a single $\pi/2$ pulse
(\ref{pi2pulse}) is
represented by a rotation of the pseudo angular momentum vector around the 
x-axis
by an angle $\pi/2$. For a single atom we have the correspondence 
$|\uparrow\rangle=|e\rangle$ and $|\downarrow\rangle=|g\rangle$. Under a 
$\pi/2$ rotation about the x-axis, the state $\uparrow\rangle$ transforms
to $|J_y=-1/2\rangle=(|\uparrow\rangle-i|\downarrow\rangle)/\sqrt{2}$ as
indicated in (\ref{pi2pulse}). For $N$ atoms, we can think of the $N$ 
individual spin ${\tiny \frac{1}{2}}$ particles combining to form a pseudo 
angular momentum 
vector of length $J=N/2$. The state of the collection of $N$ atoms can then be 
represented by appropriate superpositions of the states $|J,M\rangle$ where 
$-J\leq M\leq J$. Of course, only states within the completely symmetric 
subspace of the full $2^N$-dimensional Hilbert space can be represented in 
this way, which is justified since the coherent interaction of the 
electromagnetic fields with the atoms couple only to this symmetric subspace
(i.e. all atoms couple equally to the fields).

Free evolution in the rotating frame corresponds to
rotation of the angular momentum around the z-axis at an angular velocity
$\omega-\omega_0$. The whole Ramsey scheme can then be 
represented by the sequence: $\pi/2$ rotation about x-axis, $\phi$ rotation
about the z-axis and $-\pi/2$ rotation about the x-axis. This is the 
transformation perfomed by the unitary operator

\begin{equation}
\hat{U}(\phi)=e^{i \pi/2 \hat{J}_x}e^{-i\phi \hat{J}_z}e^{-i \pi/2
\hat{J}_x}
\end{equation}

where $\phi=(\omega-\omega_0)T$ as before.
At the end of the scheme, the number of atoms in states $|e\rangle$ and
$|g\rangle$ is measured, or equivalently their difference $\hat{J}_z(\phi)$ 
where

\begin{eqnarray}
\hat{J}_z(\phi) &=& \hat{U}(\phi)^\dagger\hat{J}_z\hat{U}(\phi) \nonumber
\\
&=& \hat{J}_z\cos\phi-\hat{J}_x\sin\phi .
\end{eqnarray}

The Ramsey signal is thus

\begin{equation}
\langle \hat{J}_z(\phi)\rangle = \langle \hat{J}_z \rangle \cos\phi - 
\langle \hat{J}_x \rangle\sin\phi
\label{jzphi}
\end{equation}

and its variance $\Delta J_z(\phi)$ is

\begin{eqnarray}
\Delta J_z(\phi) &=& \left[(\Delta J_z)^2\cos^2\phi+(\Delta J_x)^2\sin^2\phi
\right.
\nonumber \\
&-& \left. \cos\phi\sin\phi(\langle \hat{J}_x \hat{J}_z+\hat{J}_z \hat{J}_x \rangle -2 \langle \hat{J}_z \rangle 
\langle \hat{J}_x \rangle) \right]^{1/2} 
\end{eqnarray}

where the variance is defined as $(\Delta A)^2=\langle \hat{A}^2 \rangle - 
\langle \hat{A} \rangle^2$.
From the signal one wants
to estimate the phase $\phi$ and thus the frequency difference 
$\omega-\omega_0$. The phase
accuracy achievable from such a measurement is related to the signal
variance (the ``noise'') by

\begin{equation}
\delta\phi(\phi)=\frac{\Delta J_z(\phi)}{|\frac{\partial \langle 
\hat{J}_z(\phi)
\rangle}{\partial\phi}|} .
\end{equation}

For states such that $\langle \hat{J}_x \rangle =0$ (all the states we
will consider in this paper are of this type), the sensitivity 
$|\partial \langle \hat{J}_z(\phi) \rangle /\partial\phi|$ is maximal for 
$\phi=\pm\pi/2$ and the phase accuracy can be expressed as

\begin{equation}
\delta\phi(\pm\pi/2) = \frac{\Delta J_x}{|\langle \hat{J}_z \rangle |}.
\label{dphimin}
\end{equation}

Since $\Delta J_x$ and $\langle \hat{J}_z \rangle$ depend on the initial
state,
we see that different initial states lead to different phase accuracies. 
Of particular importance is the accuracy achievable when all atoms are prepared
in the same initial state. In this case the state of the atomic ensemble is a 
pure state, but it is however an uncorrelated state 
of the atomic ensemble (i.e. it can be
factorized $|\Psi\rangle = \prod_{j=1}^{N} |\psi\rangle_j$).

Consider the case of uncorrelated atoms for which all atoms have been 
prepared in the lower state $|g\rangle$, 
sometimes called a Bloch state.
The state of the atomic ensemble can thus be expressed in terms of 
eigenstates of the collective angular momentum operators as

\begin{equation}
\prod_{j=1}^{N}|g\rangle_j = |J=N/2,J_z=-N/2\rangle
\label{uncostate}
\end{equation}

where $J=N/2$ since there are $N$ $2$-level atoms, equivalent to $N$ spin 
${\tiny \frac{1}{2}}$ particles.
For such a state, the expectation value of the angular momentum operators and
their variances are calculated to be 
$\langle \hat{J}_x \rangle = \langle \hat{J}_y \rangle = 0$, $\langle \hat{J}_z \rangle = -J$,
$\Delta J_x = \Delta J_y = \sqrt{J/2}$ and $\Delta J_z = 0$. The signal and 
its variance are thus

\begin{eqnarray}
\langle \hat{J}_z(\phi) \rangle &=& -J\cos\phi \nonumber \\
\Delta J_z(\phi) &=& \sqrt{J/2}\sin\phi .
\end{eqnarray}

The maximum sensitivity is achieved at $\phi=\pm\pi/2$

\begin{equation}
\delta\phi(\pm\pi/2) = \frac{1}{\sqrt{2J}} = \frac{1}{\sqrt{N}}
\end{equation}

which is the standard quantum limit (SQL). Performing the
experiment on $N$ independent atoms all prepared in the same initial
state is thus equivalent to repeating the experiment on one atom $N$ times and
leads to an expected $1/\sqrt{N}$ factor of improvement in accuracy
over the one atom result $\Delta S_x/\langle \hat{S}_z \rangle =1$. This
is the best accuracy achievable with atoms all prepared in the same
initial pure quantum state. The number of atoms detected in the upper state, 
given by $\langle \hat{N}_+(\phi) \rangle=N/2+\langle \hat{J}_z(\phi) \rangle$,
and its variance are shown in Fig.1a.

There is a lower bound on the phase accuracy, set by Heisenberg's
uncertainty principle, $\Delta J_i \Delta J_j \geq {\tiny 
\frac{1}{2}}|\langle [\hat{J}_i,\hat{J}_j]\rangle |$ where $i,j=x,y,z$. It
is straightforward to show that 

\begin{equation}
\delta\phi\geq\frac{1}{N}
\end{equation}

which is known as the Heisenberg limit. 

We now see from (\ref{dphimin})
that in order to surpass the SQL, the atomic ensemble must be 
prepared in a state such that 
$\Delta J_x/|\langle J_z \rangle | \leq 1/\sqrt{N}$, which is a necessary and
sufficient condition for entanglement of an atomic ensemble 
\cite{squ-cold}. It is thus important to have a state for which the 
variance $\Delta J_x$ is reduced compared to its value for the uncorrelated 
state (\ref{uncostate}) while maintaining a large value for 
$\langle J_z \rangle$
so that the amplitude of the signal $\langle \hat{J}_z(\phi) \rangle =
\langle \hat{J}_z \rangle \cos\phi$ is not compromised \cite{Holland}. 
Such states which have reduced uncertainty in one observable $\Delta
J_x$ (at the expense of the conjugate observable $\Delta J_y$ having
increased fluctuations) have been called spin-squeezed states \cite{Ueda}.


\section{Spin squeezed states - Wigner function representation}

We now consider the Wigner function representation of the states 
$|\psi(a)\rangle$. The Wigner distribution of general angular-momentum states
\cite{Agarwal} is obtained from an expansion of the density operator in terms
of the multipole operators

\begin{equation}
\hat{\rho}=\sum_{k=0}^{2J}\sum_{q=-k}^{+k}\rho_{kq}\hat{T}_{kq}
\end{equation}

where the multipole operators are

\begin{eqnarray}
\hat{T}_{kq} & = & \sum_{m=-J}^{+J}\sum_{m'=-J}^{+J}(-1)^{J-m}\sqrt{2k+1}
\left(\begin{array}{ccc}J & k & J \\ -m & q & m' \end{array}\right) 
\nonumber \\
& \times & |J,m\rangle\langle J,m'|
\end{eqnarray}

and $\left(\begin{array}{ccc}J & k & J \\ -m & q & m' \end{array}\right)$ is
the usual Wigner $3j$ symbol. The wigner distribution is then given by

\begin{eqnarray}
W(\theta,\phi) &=& \sum_{k=0}^{2J}\sum_{q=-k}^{+k}Y_k^q(\theta,\phi)\rho_{kq}
\end{eqnarray}

where 
$\rho_{kq}=\langle \hat{T}_{kq} \rangle = {\rm Tr}[\hat{\rho}\hat{T}_{kq}]$ and
$Y_k^q(\theta,\phi)$ are the spherical harmonics.
In Fig. 7, the Wigner function for the state $|\psi(-1)\rangle$ clearly
shows 
the way in which this state has a large negative expectation value for 
$\hat{J}_z$, reduced variance in $\hat{J}_x$ and increased variance in 
$\hat{J}_y$. 


\begin{figure}[ht]
\begin{center}
\epsfig{file=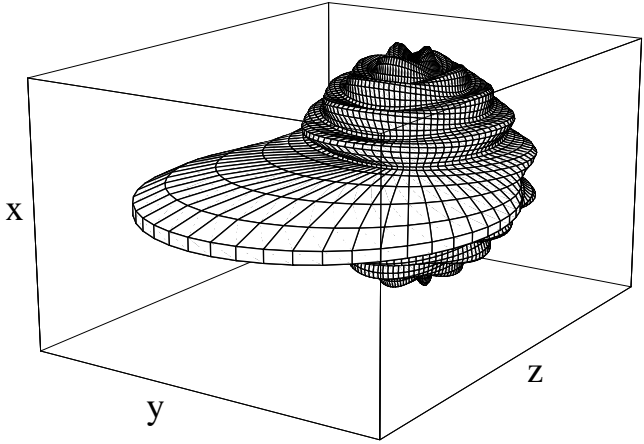,width=8.5cm}
\leavevmode
\end{center}
\vspace*{2ex}
\caption{Wigner function representation of the state $|\psi(a)\rangle$ with 
$a=-1)$. Plotted is the surface $r(\theta,\phi)=W(\theta,\phi)$, showing the
large and negative value of $\langle \hat{J}_z \rangle$, reduced variance 
$\Delta J_x$ and correspondingly increased variance $\Delta J_y$.}
\end{figure}



\section{Adiabatic elimination of excited state in Raman scattering}

From the Hamitonian (\ref{hamram}), we obtain the equations of motion for
the cavity mode and the ground state coherence $\Sigma_{gb_1}$

\begin{eqnarray}
\dot{a} &=& -\kappa a -i g_1^* \Sigma_{b_1a_1} -i g_2^*\Sigma_{ga_2}+F_a(t)
\\
\dot{\Sigma}_{gb_1} &=& -(\gamma_0-i\delta_1)\Sigma_{gb_1}+i\Omega_1
\Sigma_{a_1b_1}-i g_1^*a^\dagger \Sigma_{ga_1} \nonumber \\
&+& F_{gb_1}(t)
\label{c2}
\end{eqnarray}

and the optical polarizations associated with Stokes emission evolve according
to

\begin{eqnarray}
\dot{\Sigma}_{b_1a_1} &=& -[\gamma-i(\Delta-\delta_1)]\Sigma_{b_1a_1}
-i\Omega_1\Sigma_{b_1g}\nonumber \\
&-& i g_1 a(\Sigma_{b_1b_1}-\Sigma_{a_1a_1})+
F_{b_1a_1}(t) \\
\dot{\Sigma}_{a_1g} &=& -(\gamma+i\Delta)\Sigma_{a_1g}-i\Omega_1^*
(\Sigma_{a_1a_1}-\Sigma_{gg})\nonumber \\
&+& i g_1^*a^\dagger\Sigma_{b_1g}+F_{a_1g}(t)
\end{eqnarray}

where we assume that population in the excited state $|a_1\rangle$ decays 
towards $|b_1\rangle$ at a rate $\gamma_1$, towards $|g\rangle$ at a rate 
$\gamma_2$ and we assume a dephasing rate $\gamma_0$ for ground state 
coherences ($\gamma=(\gamma_1+\gamma_2)/2$ and $\gamma\gg\gamma_0$).

We proceed by adiabatic elimination of optical polarizations associated with
Stokes emission. To this end we assume large single-photon 
detuning $\Delta \gg \gamma$ and to first order in
$\hat{a}$ we obtain ($\Sigma_{gg}\sim N$)
\begin{eqnarray}
\Sigma_{b_1a_1} &=& \frac{\Omega_1}{\Delta}(1-i\frac{\gamma}{\Delta})
\Sigma_{b_1g}+i\frac{F_{b_1a_1}(t)}{\Delta} \\
\Sigma_{a_1g} &=& \frac{\Omega_1^*}{\Delta}N(1+i\frac{\gamma}{\Delta})-i
\frac{F_{a_1g}(t)}{\Delta}
\end{eqnarray}

which we substitute in (\ref{c2}) and obtain for the ground state spin
flip operator
$S_1=\Sigma_{gb_1}/\sqrt{N}$

\begin{eqnarray}
\dot{S_1} &=& -\left[(\gamma_0+\gamma_L)-i(\delta_1+\delta_L)\right] S_1
-i\frac{g_1^*\sqrt{N}\Omega_1}{\Delta}a^\dagger \nonumber \\
&+& {\bar F}_{S_1}(t)
\end{eqnarray}
 
where $\gamma_L=\gamma|\Omega_1|^2/\Delta^2$ is an optical pumping rate,
$\delta_L=|\Omega_1|^2/\Delta$ is the light shift and ${\bar F}_{S_1}(t)$
is a modified noise force. Light shifts can be 
incorporated in a redefinition of the energies and we ignore them in the 
remainder of this paper. Since the ground state decoherence rate is typically
very small, we also assume $\gamma_L\gg\gamma_0$ and in that limit the new
$\delta$-correlated noise forces have correlations 

\begin{eqnarray}
\langle {\bar F}_{S_1}(t){\bar F}_{S_1}^\dagger(t') \rangle &=& 2\gamma_L
\frac{\gamma_2}{\gamma}\delta(t-t') \\
\langle {\bar F}_{S_1}^\dagger(t){\bar F}_{S_1}(t') \rangle &=& 2\gamma_L
\frac{\gamma_1}{\gamma}\delta(t-t')
\end{eqnarray}


\section{Adiabatic elimination of bright polariton}

After adiabatic elimination of the excited state $|a_1\rangle$, the
relevant equations of motion are

\begin{eqnarray}
\dot{S}_1^\dagger &=& -(\gamma_L+i\delta_1)S_1^\dagger+i\chi a + 
{\bar F}_{S_1}^\dagger(t)
\nonumber \\
\dot{a} &=& -\kappa a-i\chi^*S_1^\dagger-ig_2^*\Sigma_{ga_2}+F_a(t)
\nonumber \\
\dot{S}_2 &=& -(\gamma_L+i\delta_2)S_2-i\frac{\Omega_2^*}{\sqrt{N}}
\Sigma_{ga_2}+{\bar F}_{S_2}(t)
\nonumber \\
\dot{\Sigma}_{ga_2} &=& -(\gamma+i\delta_2)\Sigma_{ga_2}-i\Omega_2\sqrt{N}S_2
-ig_2N a \nonumber \\
&+& F_{ga_2}(t)
\label{spineqns}
\end{eqnarray}

where $S_2 = \Sigma_{gb_2}/\sqrt{N}$.

From (\ref{spineqns}) and (\ref{polaritons}) and in the limit of large
ratio of 
speed of light in vacuum to group velocity of Stokes photons 
$\eta=|g_2|^2N/|\Omega_2|^2\gg1$, we obtain the equations of motion in terms
of bright and dark polaritons

\begin{eqnarray}
\dot{S}_1^\dagger &=& -(\gamma_L+i\delta_1)S_1^\dagger+
i\chi(\frac{P_D}{\eta}+P_B) +F_{S_1}^\dagger(t) 
\label{eqspinflip}
\\
\dot{P}_D &=& -(\kappa/\eta+\Gamma_2)P_D-i\frac{\chi^*}{\sqrt{\eta}}S_1^\dagger
-\frac{\kappa-\Gamma_2}{\sqrt{\eta}}P_B\nonumber \\
&+& F_D(t)
\label{eqdarkpol}
\\
\dot{P}_B &=& -(\kappa+\Gamma_2/\eta)P_B-i\chi^*S_1^\dagger-ig_2\sqrt{N}
{\tilde \Sigma}_{ga_2}\nonumber \\
&+&F_B(t)
\label{eqpolb} \\
\dot{{\tilde \Sigma}}_{ga_2} &=& -\Gamma{\tilde \Sigma}_{ga_2}-ig_2\sqrt{N}P_B
+{\tilde F}_{ga_2}(t)
\label{eqoptcoh}
\end{eqnarray}

where $\Gamma_2=\gamma_L+i\delta_2$, $\Gamma=\gamma+i\delta_2$ and 
${\tilde \Sigma}_{ga_2}=\Sigma_{ga_2}/\sqrt{N}$ and noise forces were
modified
appropriately. Note that in the picture of dark and bright polaritons,
only
the bright polariton is coupled to the excited state through the optical
coherence $\Sigma_{ga_2}$.

Under adiabatic conditions, the bright polariton evolves slowly (on a typical
timescale $T$) and we can solve perturbatively in $1/T$. The equations 
(\ref{eqpolb}) and (\ref{eqoptcoh}) are of the form $\dot{{\bf x}}=
-{\bf M}.{\bf x}+{\bf y}$, where ${\bf x}$ is the vector $(P_B,{\tilde
\Sigma}_{ga_2})$, ${\bf M}$ is a $2\times 2$ matrix and ${\bf y}$ is a 
source term 

\begin{eqnarray}
\frac{d}{dt}
\left[\begin{array}{c}P_B \\ {\tilde \Sigma}_{ga_2}\end{array}\right]
&=&-\left(\begin{array}{cc}\kappa & ig_2\sqrt{N} \\ ig_2\sqrt{N} &
\Gamma\end{array}\right)
\left[\begin{array}{c}P_B \\ {\tilde \Sigma}_{ga_2}\end{array}\right]
\nonumber \\
&+&
\left[\begin{array}{c}-i\chi^*S_1^\dagger-\frac{\kappa}{\eta}P_D+F_B(t) \\ 
{\tilde F}_{ga_2}(t)\end{array}\right]
\label{mateqn}
\end{eqnarray}

where we have used $\kappa\gg\gamma_L$ and where $F_B(t)$ and ${\tilde
F}_{ga_2}(t)$ are appropriate noise forces. 
These equations can be solved 
easily to first order by ${\bf x}^{(0)}(t)={\bf M}^{-1}.{\bf y}$, higher
order approximations yielding ${\bf x}^{(n)}(t)={\bf M}^{-1}.[{\bf y}-
\dot{{\bf x}}^{(n-1)}(t)]$.

We can rewrite 
\begin{equation}
\frac{|g_2|^2N}{\kappa\gamma} \sim 3\pi\times\left(\frac{N}{V}L\lambda^2\right)
\times{\mathcal F}
\end{equation}

i.e. the density length product multiplied by the cavity finesse, so 
that with densities corresponding to room temperature atomic vapours, optical
wavelengths and finesse of order $100$ this quantity is already of order  
$\sim 10^4$.
We can thus assume that $|g_2|^2N/(\kappa\gamma)\gg 1$ and solve in powers
of $\kappa\gamma/(|g_2|^2N)$. 

We see from (\ref{mateqn}) that ${\bf x}^{(n)}(t)$ is of order
$[\kappa\gamma/(|g_2|^2N)]^{(n+1)}$ and thus solving to lowest order we
find

\begin{eqnarray}
P_B &=& \frac{1}{|g_2|^2N}[-i\Gamma\chi^*S_1^\dagger-
\frac{\kappa\Gamma}{\sqrt{\eta}}P_D+\Gamma F_B(t) \nonumber \\
&-& ig_2F_{ga_2}(t)]
\end{eqnarray}

so that when $\eta\gg 1$,

\begin{eqnarray}
a &\simeq& \frac{P_D}{\eta}+P_B \nonumber \\
&\simeq& \frac{P_D}{\eta}+\frac{1}{|g_2|^2N}[-i\Gamma\chi^*S_1^\dagger
+\Gamma F_B(t) \nonumber \\
&-& ig_2F_{ga_2}(t)].
\end{eqnarray}

The coupled equations of motion for the dark state polariton (\ref{eqdarkpol})
and the spin flip (\ref{eqspinflip}) then become

\begin{eqnarray}
\dot{S}_1^\dagger &=& (\frac{|g_1|^2}{|g_2|^2}\gamma_L-\gamma_L-i\delta_1)
S_1^\dagger + i\frac{\chi}{\sqrt{\eta}}P_D +{\tilde F}_{S_1}^\dagger(t) 
\label{eqfinspin}
\\
\dot{P}_D &=& -(\kappa/\eta+\gamma_L+i\delta_2)P_D-i\frac{\chi^*}{\sqrt{\eta}}
S_1^\dagger + {\tilde F}_D(t)
\label{eqfinpol}
\end{eqnarray}

where ${\tilde F}_{S_1}^\dagger(t)$ and ${\tilde F}_D(t)$ are modified noise
forces with correlations

\begin{eqnarray}
\langle {\tilde F}_D(t){\tilde F}_D^\dagger(t') \rangle &=& \frac{\kappa}{\eta}
+2\gamma_L\frac{\gamma_2}{\gamma} \\
\langle {\tilde F}_D^\dagger(t){\tilde F}_D(t') \rangle &=& 2\gamma_L\frac{\gamma_1}{\gamma} \\
\langle {\tilde F}_{S_1}(t){\tilde F}_{S_1}^\dagger(t') \rangle &=& 2\gamma_L\frac{\gamma_2}{\gamma} \\
\langle {\tilde F}_{S_1}^\dagger(t){\tilde F}_{S_1}(t') \rangle &=& 2\gamma_L\frac{\gamma_1}{\gamma} \\
\end{eqnarray}

and all other correlations can be neglected. The coherent part of the 
interaction can thus be obtained from an effective hamiltonian 

\begin{equation}
H_{eff} = \frac{\hbar\chi}{\sqrt{\eta}}S_1P_D+{\rm h.c.}
\end{equation}

where the interaction rate is $\chi/\sqrt{\eta}=g_1\Omega_1^*|\Omega_2|/(|g_2|
\Delta)$.

We note (\ref{eqfinpol}) that cavity losses are strongly suppressed in the 
limit $\eta \gg 1$.
Indeed, subsequent to the large group velocity reduction \cite{slow-light}, 
the polariton is almost purely atomic and the excitation leaks very slowly 
out of the medium. 
The equation of motion for coherence $S_1^+$ (\ref{eqfinspin}) contains 
a loss term 
(due to isotropic spontaneous emission) and a linear gain term (due to emission
into bright polariton). 
The two can compensate each other. However the linear phase-insensitive 
amplification is also accompanied by
correspondingly increased fluctuations, represented by new Langevin forces 
$\tilde{F}_D(t),\tilde{F}_{S_1}^+(t)$. 
In the case that $g_1=g_2$ and when all Rabi frequencies are taken to be
real, we have the interaction rate
$\xi=\chi/\sqrt{\eta}=\Omega_1\Omega_2/\Delta$.


\def\etal{\textit{et al.}}

\end{document}